\documentstyle[fancyheadings,prl,
amsmath,amsfonts,isolatin1,aps,prc,epsf,amssymb,epsfig,multicol]{revtex}

\def\be{\begin{equation}}
\def\ee{\end{equation}}
\def\bea{\begin{eqnarray}}
\def\eea{\end{eqnarray}}
\def\bea{\begin{eqnarray*}}
\def\eea{\end{eqnarray*}}
\def\nn{\nonumber}

\begin{document}

\sloppy

\draft

\bibliographystyle{srt}

\title{ Preheating in Quintessential Inflation}

\author{ A. H. Campos$^{a,}$\footnote{hcampos@charme.if.usp.br},
         H. C. Reis$^{b,}$\footnote{hreis@ifi.unicamp.br},
         R. Rosenfeld$^{c,}$\footnote{rosenfel@ift.unesp.br}}

\address{\it  $^a$ Instituto de F\'{\i}sica - USP}
\address{\it Departamento de F\'{\i}sica Nuclear}
\address{\it CP 66318, S\~{a}o Paulo, SP, Brazil}
\vspace{1cm}

\address{\it  $^b$ Instituto de F\'{\i}sica Gleb Wataghin - UNICAMP}
\address{\it Departamento de Raios C\'osmicos e Cronologia}
\address{\it CP 6165, Campinas, SP, Brazil}
\vspace{1cm}

\address{\it  $^c$ Instituto de F\'{\i}sica Te\'orica - UNESP}
\address{\it Rua Pamplona, 145}
\address{\it 01405-900 S\~{a}o Paulo, SP, Brazil}

\maketitle

\vspace{0.1cm}

\begin{abstract}
We perform a numerical study of the preheating mechanism of particle
production in models of
quintessential inflation and compare it
with the usual gravitational production mechanism. We find that even for a
very small coupling between the inflaton field and a massless scalar field,
$g \gtrsim 10^{-9}$, preheating dominates over
gravitational particle production. Reheating temperatures in the range
$10^{-2}\;\;\mbox{GeV} \lesssim T_{rh} \lesssim 10^{15} \;\; \mbox{GeV}$
can be easily obtained.
\end{abstract}

\vspace{0.2cm}

\noindent
PACS Categories:    98.80.Cq

\vspace{0.2cm}


\begin{multicols}{2}
\relax

\section{Introduction}

Inflationary models of the early universe have become a paradigm that can
explain the isotropy of the cosmic microwave background, the origin of the
small adiabatic perturbations that seed galaxy formation and the flatness of
space-time. There is also evidence that the universe has recently entered
another stage of mild inflation, responsible for its accelerated expansion.
It would be natural to investigate models where the two periods of inflation
are related. In
particular, models of quintessential inflation explore the possibility that
the same scalar field is responsible for both periods of inflation
\cite{Peebles-Vilenkin,Spokoiny}.

In quintessential inflation models, the inflaton potential has no minimum in
which the field could oscillate to produce the matter that would reheat the
universe, as in usual chaotic inflation models. Therefore, it is assumed that
all matter and energy in the universe will be created by gravitational
production associated to changes in the geometry of the universe.
However, it is known that this process is not very efficient \cite{Ford}.
In this letter we assess the importance of another mechanism of particle
production in the early universe, namely preheating \cite{preheating}, in which
particles are produced due to the variation of the classical inflaton
field.
We find that the introduction of a non-zero coupling between
the inflaton and matter fields leads to particle production that can easily
overcome the gravitational one.


\section{The model}

For definiteness, we study the Peebles and Vilenkin model \cite{Peebles-Vilenkin},
with a potential for the inflaton field $\Phi$ of the form
\bea
V(\Phi) & = & \lambda (\Phi^4+M^4) \quad \mbox{for} \quad \Phi < 0, \nn \\
  &   & \frac{\lambda M^8}{\Phi^4 + M^4} \quad \mbox{for} \quad \Phi \geq  0,
\eea
where $\lambda = 1 \times 10^{-14}$ is required from structure formation
\cite{Linde}
and $M = 5 \times 10^8$ GeV in order to produce the present quintessential
energy density \cite{Peebles-Vilenkin}.

Defining adimensional parameters $\phi = \Phi/M_{Pl}$, $q=M/M_{Pl}$ and
$\tau = \sqrt{\lambda} M_{Pl} t$, the classical equation of motion for
the inflaton field becomes independent of the $\lambda$ parameter:

\begin{eqnarray}
\phi '' &+& \sqrt{24 \pi} \sqrt{\frac{\phi^{'2}}{2}  +
(\phi^4+q^4)} \;
\phi' + \\ \nonumber
&&  4 \phi^3 = 0 \;\;\; \mbox{if} \;\; \phi <0, \\
\phi '' &+& \sqrt{24 \pi} \sqrt{\frac{\phi^{ '2}}{2}  + \frac{q^8}
{(\phi^4+q^4)}} \;
\phi' - \\ \nonumber
&&4 \frac{q^8}
{(\phi^4+q^4)^2} \phi^3 = 0 \;\;\; \mbox{if} \;\; \phi \geq 0,
\label{eqphi}
\end{eqnarray}
where  the primes are the derivatives with respect to $\tau$ and
we have used the Hubble parameter,
\begin{equation}
H^2(\tau) = \left(\frac{a'(\tau)}{a(\tau)} \right)^2 =
\frac{8\pi }{3}  \left( \frac{\phi^{'2}}{2}
 + V(\phi) \right).
\label{hubble-phi}
\end{equation}

In order to study preheating, we couple the classical inflaton field to a
massless quantum scalar field $\chi$ through the lagrangian:
\begin{equation}
{\cal L}=
\frac{1}{2} \partial_\mu \chi \partial^\mu \chi -
\frac{1}{2} \xi R \chi^2
-\frac{1}{2} g^2\Phi^2\chi^2 ,
\label{lagrangeana}
\end{equation}
where $R$ is the scalar curvature, $\xi$ is the gravitational coupling and
$g$ is a quartic
coupling constant between $\phi$ and $\chi$. Ford \cite{Ford} calculated the
energy density
of $\chi$ particles produced gravitationally in the limit $|\xi - 1/6| \ll
1$ due to the transition of the universe from the de Sitter space-time at
the end of inflation. However, this production is not efficient since the ratio
of matter and inflaton field energy densities is typically
$\rho_m/\rho_\phi \sim 10^{-18}$ \cite{Ford}. After thermalization of the
products it is possible to define a
reheating temperature when  radiation starts to dominate and in that case
it is of the order of $T_{rh} \sim 10^4$ GeV.

When preheating is operative, the picture changes drastically.
Taking into account the dilution of gravitational
particle production, at the end of inflation, until the beginning of the
particle production by the
preheating mechanism, the universe is still dominated by the inflaton density
energy and this is our starting point.

The equation of motion for each comoving $k$ mode of the $\chi$ field can be
written in
terms of a new convenient variable $X_k =a^{3/2}\chi_k $ as:
\begin{equation}
X_k '' + \left\{ \frac{1}{\lambda} \left( \frac{k^2}{M_{Pl}^2 a^2(\tau)} + g^2
\phi^2(\tau)\right) - \Delta \right\}  X_k =0.
\label{eq.movimento-X}
\end{equation}
The physical momentum is given by $k/a$  and we have verified
that $\Delta=\frac{3}{4} \left( \frac{a'}{a} \right) ^2 +
\frac{3}{2} \left(\frac{a''}{a}\right)$ is neglegible during the production of
$\chi$ particles.

This is an equation of motion for an harmonic oscillator
with variable frequency,
$\omega_k^2=(k^2/(M_{Pl}^2 a^2(\tau))+g^2 \phi^2(\tau))/\lambda$.
When the adiabaticity condition is broken, that is, $\omega_k' \geq \omega_k^2$,
particles in the $k$ mode are produced.
Figure 1 illustrates
the time interval where adiabaticity is violated, leading to
particle creation. This time interval corresponds to the passage of the
inflaton field through its origin.

\begin{figure}[th]
\vspace*{5mm}
\centerline{\epsfxsize=3.5in\epsfbox{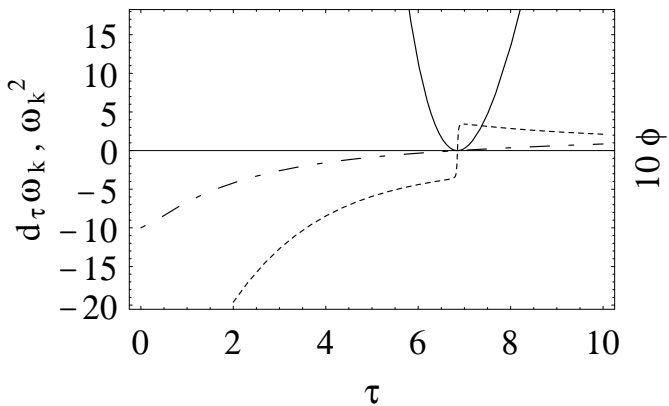}}
\vspace*{9mm}
\caption{
$\omega_k^2$  (solid line), $\omega_k'$ (dotted line) and $10 \times \phi$
(dot-dashed line) as a function of $\tau$
for $g=10^{-5}$ and $k=5 \sqrt\lambda M_{Pl}^2$. Notice that adiabaticity is
violated around $\tau \simeq 7$, when the inflaton field goes through zero.}
\label{adiabaticity}
\end{figure}

\section{Particle production}

\begin{figure}[th]
\vspace*{5mm}
\centerline{\epsfxsize=3.5in\epsfbox{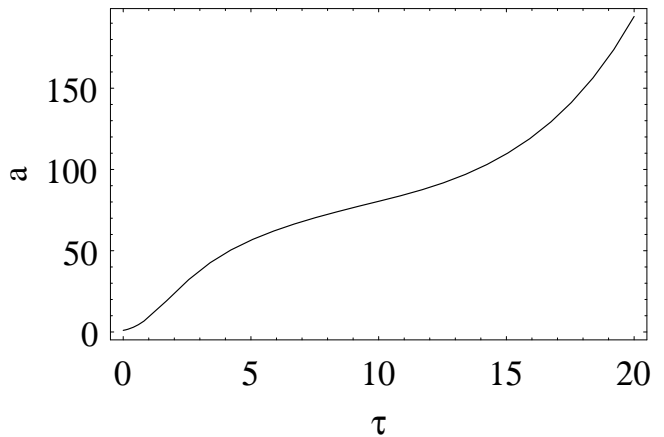}}
\vspace*{9mm}
\caption{Evolution of the scale factor
as a function of $\tau$.}
\label{scalefactor}
\end{figure}

In order to compute the total amount of particles produced during preheating, we
have to solve equation (\ref{eq.movimento-X}). To obtain $\phi(\tau)$ we choose
the initial conditions as being $\phi(\tau=0)=-1$ and
$\phi'(\tau=0)=0$, since in $\lambda \Phi^4$ model inflation ends
when $\Phi \simeq - M_{Pl}$. Then we can evaluate the
evolution of the scale factor $a(\tau)$ given by:
\begin{equation}
a(\tau) = a_0 \exp \left[  \int_0^\tau \; d\tau' H(\tau') \right],
\label{scale-factor}
\end{equation}
where $H(\tau)$ is given in equation (\ref{hubble-phi}) and $a_0 = 1$.
We show in figure \ref{scalefactor} the evolution of the scale factor during the
period relevant for particle production.

Finally, substituting (\ref{scale-factor}) and the solution of equations
($1-2$) into
(\ref{eq.movimento-X}) we obtain $X_k(\tau)$.

We define an adiabatic invariant $n_k$ that can be interpreted as the
comoving number density of particles produced with momentum $k$:
\begin{equation}
n_k= \frac{\omega_k}{2} \left(\frac{|X_k'|^2}{\omega_k^2} +|X_k|^2 \right) -
      \frac{1}{2}.
\label{number}
\end{equation}

\begin{figure}[th]
\vspace*{5mm}
\centerline{\epsfxsize=3.5in\epsfbox{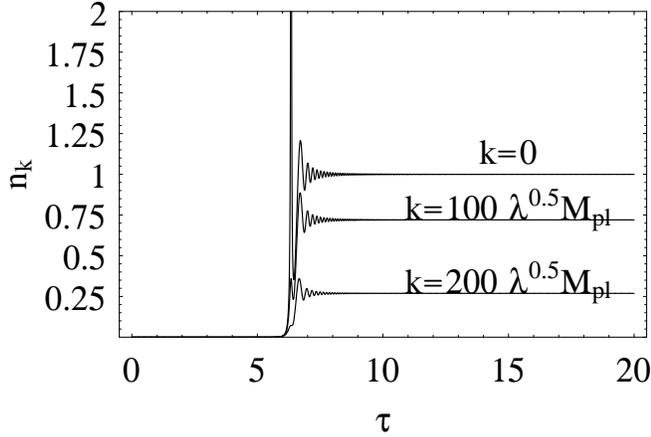}}
\vspace*{9mm}
\caption{Comoving number density of $\chi$ particles produced during preheating
as a function of $\tau$ for comoving momentum
$k= 0, 100$ and $200$ $\sqrt{\lambda} M_{Pl}$, for
$g = 5 \times 10^{-5}$. }
\label{numberK}
\end{figure}

In figure \ref{numberK} we show the comoving number density of particles
created during preheating obtained from the solution of equation
(\ref{eq.movimento-X})
with vacuum initial conditions for different values of comoving momentum $k$, for
$g = 5 \times 10^{-5}$. One can see that particle production occurs at $\tau
\simeq 7$, and quickly stabilizes.

\begin{figure}[th]
\vspace*{5mm}
\centerline{\epsfxsize=3.5in\epsfbox{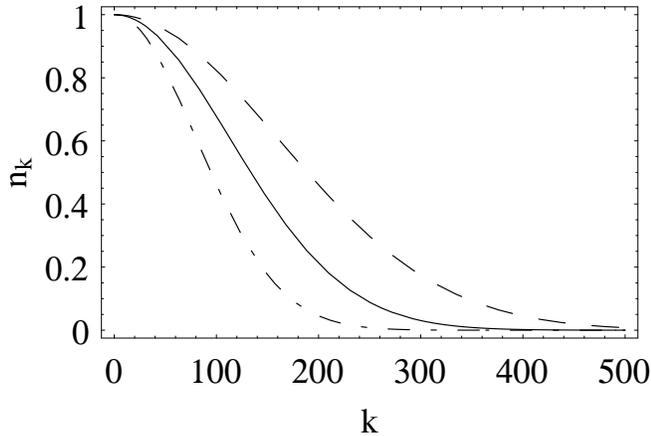}}
\vspace*{9mm}
\caption{Comoving number density of $\chi$ particles produced during preheating
as a function of comoving momentum $k$ in units of $\sqrt{\lambda} M_p$.
$\lambda = 1 \times 10^{-14}$ and $g = 5 \times 10^{-5}$ (solid curve);
$\lambda = 1 \times 10^{-14}$ and $g = 1 \times 10^{-4}$ (dashed curve);
$\lambda = 4 \times 10^{-14}$ and $g = 5 \times 10^{-5}$ (dot-dashed curve). }
\label{spectrum}
\end{figure}

In order to illustrate the $k-$ dependence of the density of particles, we plot
in figure \ref{spectrum} the density spectrum
produced during preheating for different values of $\lambda$ and $g$.

The spectrum is well described by the expression:
\begin{equation}
n_k = \exp \left[ -\frac{1}{52} \frac{\sqrt{\lambda}}{g} \frac{k^2}{\lambda
M_{Pl}^2} \right].
\end{equation}
This result is to be compared to the approximation for particle production in
the first oscillation of a pure $\phi^4$ potential \cite{FKL}:
\begin{equation}
n_k^{appr.} = \exp \left[ - \pi \frac{\lambda M_{Pl}^2 }{g
|\dot{\Phi}_p|} \frac{k^2/a^2(\tau_p)}{\lambda M_{Pl}^2} \right],
\end{equation}
where $\tau_p$ is the time of particle production and $|\dot{\Phi}_p|$ is the velocity
of the inflaton field at the bottom of the potential. In our case we
have $|\dot{\Phi}_p| = 3.6 \times 10^{-2} \sqrt{\lambda} M_{Pl}^2$ and
$a(\tau_p) = 68$, which results in:
\begin{equation}
n_k^{appr.} = \exp \left[ -\frac{1}{53} \frac{\sqrt{\lambda}}{g} \frac{k^2}{\lambda
M_{Pl}^2} \right] ,
\end{equation}
an excellent approximation indeed.

The total energy density produced at $\tau_p$ is given by:
\begin{equation}
\rho_\chi =  \int \frac{d^3 k}{(2 \pi )^3 a^3(\tau_p)} \frac{k}{a(\tau_p)} \; n_k \simeq
3 \times 10^{-6} \lambda g^2 M_{Pl}^4
\label{X-production}
\end{equation}

Since the energy density in the inflaton field around the time of particle
production is
\begin{equation}
\rho_\phi \simeq \frac{\dot{\Phi}_p^2}{2} \simeq 6 \times 10^{-4} \lambda
M_{Pl}^4 \; ,
\end{equation}
the energy density that is transferred from the inflaton field to the
$\chi$ field at the moment of production is given by:
\begin{equation}
\frac{\rho_\chi}{\rho_\phi} \simeq 5 \times 10^{-3} g^{2}
\end{equation}
This means that there is an upper limit in energy that can be
extracted
from $\phi$ field, at the moment of production,
$\frac{\rho_\chi}{\rho_\phi} \simeq 5 \times 10^{-3}$.
However, once the $\chi$-particles are produced they rapidly become
non-relativistic, since their effective masses are given by $m_{\chi}= g\Phi$
(see equation(\ref{eq.movimento-X})). In such a case, the above comparison
becomes
\begin{equation}
\frac{\rho_\chi}{\rho_\phi} \simeq 14 g^{5/2},
\end{equation}
for $\Phi \simeq 10^{-1}M_{Pl}$ which is a
good approximation for early times as shown in figure 1.

In order to prove the initial statement that
the diluted energy density of particles produced gravitationally, $\rho_m$
\cite{Ford} is in fact much smaller than the inflaton energy density, $\rho_\phi$ at $\phi \sim
0$, when the preheating starts, we take the ratio between them
\footnote{Although $\rho_\phi$ is not proportional to $a^{-6}$ through all the
interval, it corresponds to a maximum dilution.}:
\begin{equation}
\frac{\rho_m(\phi=0)}{\rho_\phi(\phi=0)} \sim \frac{\rho_m(\phi=-1)/a^{-4}}{\rho_\phi(\phi=-1)/a^{-6}}
\sim 10^{-22}
\end{equation}
On the other hand, we can compare $\rho_m$ with our $\chi$ production (equation
(\ref{X-production})):
\begin{equation}
\frac{\rho_{\chi}(\phi=0)}{\rho_m(\phi=0)} \sim 10^{23} g^{5/2}
\end{equation}
and conclude that the latter dominates over the former for
$g\gtrsim 6 \times 10^{-10}$.

Finally, let us estimate the reheating temperature. We will make the assumption
that the $\chi$ particles decay fast enough to avoid the effects of backreaction
over the $\phi$ field. In reference \cite{FKL} it is shown that backreaction is
important only for
large times. So, without loss of generality, we will consider the transference
of $\chi$ energy to relativistic particles just after the $\chi$ production.

As the inflaton field energy density scales as $a^{-6}$ since it is dominated by
kinetic energy whereas the relativistic particles energy
density, $\rho_R$, scales as $a^{-4}$, the latter will start to dominate the
universe at a time $t_d$ given
by:
\begin{equation}
\frac{a^2(\tau_p)}{a^2(\tau_d)} \simeq 14 g^{5/2}
\end{equation}

At this time, the $\rho_R$ will be diluted to the value:
\bea
\rho_R & = & 3 \times 10^{-5} \; \lambda^{3/4} g^{5/2} M_{Pl}^3 \;\Phi \;\frac{a^4(\tau_p)}{a^4(\tau_d)}
\nonumber \\
       & \simeq & 6 \times 10^{-4} \lambda^{3/4} g^{15/2} M_{Pl}^4,
\eea
and the reheating temperature will be:
\begin{equation}
 T_{rh}  \sim  (\rho_R)^{1/4} \Longrightarrow
T_{rh} \simeq 10^{-1} \;(\lambda^{3/4} g^{15/2})^{1/4} \; M_{Pl} .
\end{equation}

For our limits on $g$, $6 \times 10^{-10}\lesssim g \lesssim 1$,
the reheating temperature corresponds respectively to:
\begin{equation}
10^{-2}\;\;\mbox{GeV} \lesssim T_{rh} \lesssim 10^{15} \;\; \mbox{GeV},
\end{equation}
which is a confortable range.
Further constraints in $T_{rh}$ imply in new bounds on the
coupling constant between the inflaton and the $\chi$ fields.

\section{Conclusions}

Quintessential inflation models postulate that the same field is responsible for
inflation in the early universe and the accelerated expansion of the universe
today. In this type of models, the potential has no minimum and the traditional
reheating mechanism of the universe can not be operative. Gravitational
production of particles is usually adopted to reheat the universe.

We have demonstrated in this letter that introducing a small coupling
$g\gtrsim 10^{-9}$
between the inflaton field and a massless scalar particle $\chi$ leads to
particle creation in the preheating mechanism that dominates over the usual
gravitational production in quintessential inflation models.

Reheating temperatures in the range
$10^{-2}\;\;\mbox{GeV} \lesssim T_{rh} \lesssim 10^{15} \;\; \mbox{GeV}$
can be easily obtained for $6 \times 10^{-10}\lesssim g \lesssim 1$.

Our result is
stronger than the one obtained in the approximation used in \cite{FKL} because
we took into account the dilution of the produced particles between the epochs
of gravitational creation and preheating. It is also not fully complete
since the gravitational production mechanism could be affected when the
$\chi-\phi$ coupling is
introduced in the lagrangian. However, the effect would be to inhibit the
gravitational production, making our results even stronger.

\section*{Acknowledgments}

The authors would like to thank Fapesp and CNPq for partial
financial support.

\def \arnps#1#2#3{Ann.\ Rev.\ Nucl.\ Part.\ Sci.\ {\bf#1} (#3) #2}
\def \art{and references therein}
\def \cmts#1#2#3{Comments on Nucl.\ Part.\ Phys.\ {\bf#1} (#3) #2}
\def \cn{Collaboration}
\def \cp89{{\it CP Violation,} edited by C. Jarlskog (World Scientific,
Singapore, 1989)}
\def \econf#1#2#3{Electronic Conference Proceedings {\bf#1}, #2 (#3)}
\def \efi{Enrico Fermi Institute Report No.\ }
\def \epjc#1#2#3{Eur.\ Phys.\ J. C {\bf#1} (#3) #2}
\def \f79{{\it Proceedings of the 1979 International Symposium on Lepton and
Photon Interactions at High Energies,} Fermilab, August 23-29, 1979, ed. by
T. B. W. Kirk and H. D. I. Abarbanel (Fermi National Accelerator Laboratory,
Batavia, IL, 1979}
\def \hb87{{\it Proceeding of the 1987 International Symposium on Lepton and
Photon Interactions at High Energies,} Hamburg, 1987, ed. by W. Bartel
and R. R\"uckl (Nucl.\ Phys.\ B, Proc.\ Suppl., vol.\ 3) (North-Holland,
Amsterdam, 1988)}
\def \ib{{\it ibid.}~}
\def \ibj#1#2#3{~{\bf#1} (#3) #2}
\def \ichep72{{\it Proceedings of the XVI International Conference on High
Energy Physics}, Chicago and Batavia, Illinois, Sept. 6 -- 13, 1972,
edited by J. D. Jackson, A. Roberts, and R. Donaldson (Fermilab, Batavia,
IL, 1972)}
\def \ijmpa#1#2#3{Int.\ J.\ Mod.\ Phys.\ A {\bf#1} (#3) #2}
\def \ite{{\it et al.}}
\def \jhep#1#2#3{JHEP {\bf#1} (#3) #2}
\def \jpb#1#2#3{J.\ Phys.\ B {\bf#1} (#3) #2}
\def \jpg#1#2#3{J.\ Phys.\ G {\bf#1} (#3) #2}
\def \mpla#1#2#3{Mod.\ Phys.\ Lett.\ A {\bf#1} (#3) #2}
\def \nat#1#2#3{Nature {\bf#1} (#3) #2}
\def \nc#1#2#3{Nuovo Cim.\ {\bf#1} (#3) #2}
\def \nima#1#2#3{Nucl.\ Instr.\ Meth. A {\bf#1} (#3) #2}
\def \npb#1#2#3{Nucl.\ Phys.\ B {\bf#1} (#3) #2}
\def \npps#1#2#3{Nucl.\ Phys.\ Proc.\ Suppl.\ {\bf#1} (#3) #2}
\def \npbps#1#2#3{Nucl.\ Phys.\ B Proc.\ Suppl.\ {\bf#1} (#3) #2}
\def \PDG{Particle Data Group, D. E. Groom \ite, \epjc{15}{1}{2000}}
\def \pisma#1#2#3#4{Pis'ma Zh.\ Eksp.\ Teor.\ Fiz.\ {\bf#1} (#3) #2 [JETP
Lett.\ {\bf#1} (#3) #4]}
\def \pl#1#2#3{Phys.\ Lett.\ {\bf#1} (#3) #2}
\def \pla#1#2#3{Phys.\ Lett.\ A {\bf#1} (#3) #2}
\def \plb#1#2#3{Phys.\ Lett.\ B {\bf#1} (#3) #2}
\def \pr#1#2#3{Phys.\ Rev.\ {\bf#1} (#3) #2}
\def \prc#1#2#3{Phys.\ Rev.\ C {\bf#1} (#3) #2}
\def \prd#1#2#3{Phys.\ Rev.\ D {\bf#1} (#3) #2}
\def \prl#1#2#3{Phys.\ Rev.\ Lett.\ {\bf#1} (#3) #2}
\def \prp#1#2#3{Phys.\ Rep.\ {\bf#1} (#3) #2}
\def \ptp#1#2#3{Prog.\ Theor.\ Phys.\ {\bf#1} (#3) #2}
\def \ppn#1#2#3{Prog.\ Part.\ Nucl.\ Phys.\ {\bf#1} (#3) #2}
\def \rmp#1#2#3{Rev.\ Mod.\ Phys.\ {\bf#1} (#3) #2}
\def \rp#1{~~~~~\ldots\ldots{\rm rp~}{#1}~~~~~}
\def \si90{25th International Conference on High Energy Physics, Singapore,
Aug. 2-8, 1990}
\def \zpc#1#2#3{Zeit.\ Phys.\ C {\bf#1} (#3) #2}
\def \zpd#1#2#3{Zeit.\ Phys.\ D {\bf#1} (#3) #2}

\end{multicols}

\end{document}